\documentclass[12pt,a4paper,twoside]{article}        
\usepackage[]{babel}                          
\usepackage[cp1251]{inputenc}                        
\usepackage{mmgart}                                  

\begin{document}

\setcounter{page}{13}                                
\thispagestyle{empty}                                
\begin{heading}                                      
{Volume\;9,\, N{o}\;1,\, p.\,1 -- 11\, (2021)}      
{}
\end{heading}                                        

\begin{Title}
Particle configurations in the $NN\bar K$  system
\end{Title}

\begin{center}
\Author{1,a}{Igor Filikhin},
\Author{2}{Yury B. Kuzmichev}
\and
\Author{1}{Branislav Vlahovic}
\end{center}


\begin{flushleft}
\Address{1}{Department of Mathematics and Physics, North Carolina Central University, %
       Durham, NC 27707, USA}

\Address{2}{Yaroslavl State Pedagogical University, 150000, Yaroslavl, Russia}


\Email{$^a$\,ifilikhin@nccu.edu}
\end{flushleft}

\Headers{I. Filikhin et al}{Particle configurations in $NN\bar K$  system}

\begin{flushleft}                                 
\end{flushleft}                                   

\Thanks{This work is supported by the National Science Foundation grant HRD-1345219 and DMR-1523617 awards
Department of Energy$/$National Nuclear Security Administration under Award Number NA0003979
DOD-ARO grant W911NF-13-0165.}

\Thanks{\mbox{}\\
\copyright\,The author(s) 2021. \ Published by Tver State University, Tver, Russia}
\renewcommand{\thefootnote}{\arabic{footnote}}
\setcounter{footnote}{0}

\Abstract{Three-body $AAB$ model for the $NN{\bar K}(s_{NN}=0)$ kaonic cluster is considered based on the configuration space Faddeev equations.
Within a single-channel approach, the difference between masses of nucleons and kaons and the charge independence breaking of nucleon-nucleon interaction are taken into consideration.
We definite the particle configurations in the system according to the particle masses and pair potentials. There are two sets of the particle configurations, $ ppK^-$, $np \bar {K^0}$ and $nn{\bar K}^0$, $npK^-$, charged and neutral.
The three-body calculations are performed by applying $NN$ and $N\bar K$ phenomenological isospin-dependent potentials. The mass and energy spectra related to the particle configurations are presented. We evaluate the mass and energy uncertainties for the $NN\bar K$ model.  An analogy to  $NNN$ model for the $^3$H and $^3$He nuclei is proposed.}




\newpage                               
\renewcommand{\baselinestretch}{1.1}   


\section{Introduction}
The quasi-bound state of the kaonic cluster "$ppK^-$" \cite{AY02}
has been intensively debated during the last time.
The theoretical predictions for the binding energy for this $NN{\bar K}(s_{NN}=0)$ system obtained within different models demonstrated significant disagreement with the values derived from early experimental data \cite{G2016}. The experimentally motivated value was proposed to be about -100~MeV (deeply bound state) \cite{Y10,I15}. The discussions about the experimental background and theoretical interpretations can be found in Refs. \cite{GHM,H18,SOR18,HAY2017,AY19,S2017}.

The newest J-PARC E15 experiments \cite{J2019,SOR19,Y20} had reported the value between -43 and -47~MeV.
This value is close to the theoretical prediction made within the phenomenological approach of Ref. \cite{YA07}.
The numerical analysis \cite{VF20a,FV20b} shows that there is
a lower bound of the ''$ppK^-$ energy for the phenomenological potentials having the Gaussian form. The lower bound is about -60~MeV for the $N{\bar K}$ potential proposed in Ref. \cite{YA07}.
This potential has a significant difference for the isospin singlet and triplet channels.
The isospin singlet component of the ${\bar K}N$ potential generates a quasi-bound state corresponding to the $\Lambda(1405)$ resonance below the $K^-p$ threshold.
The discussed is the nature of the ${\bar K}N$ interaction, which would include two channels
$N{\bar K}$ quasi-bound state and $\pi\Sigma$ resonance \cite{YA07,D1960,J10,170,171}.
To treat this complication, the
Akaishi-Yamazaki (AY) $N{\bar K}$ potential \cite{YA07} takes into account the $\pi\Sigma$ coupling effectively.
This effective
${\bar K}N$ interactions have a strong attraction in the singlet $I=0$ channel
and a weak attraction in the triplet $I=1$ channel. The two-body threshold
is close to the bound state energy of $\Lambda(1405)$ (about -27~MeV) as the $K^-p$ bound pair.
The binding energy of $ppK^-$ obtained within $NN{\bar K}(s_{NN}=0)$ isospin model is about -48~MeV \cite{YA07}.
Similar results have been obtained within different models in Refs. \cite{D17,MKE16,RS14,S20}.

In this work, we study the "$ppK^-$" kaonic cluster within the isotopic-spin formalism.
Recently, the isospin model for $NN{\bar K}$ system has been presented in the terms of the particle sets. In particular,
the particle channels $ppK^-$ and $pn{\bar K}^0$ have been defined in Refs. \cite{O17,H17} for the "$ppK^-$" cluster due to the possible particle transition $n{\bar K}^0\longleftrightarrow pK^-$. The channels are coupled. Thus, the system may be found as $ppK^-$ or $pn{\bar K}^0$ at the same time with diffident projections of total isospin (I=1/2). Such consideration allows to definite a particle model for the $NN{\bar K}(s_{NN}=1)$ cluster \cite{R16}.
This "particle representation" is essentially based on the "charge" isospin basis \cite{D15}, which can be obtained by unitary rotation from a "natural" isospin basis \cite{FV20}. Both formulations are equivalent. However, the last one does not support channel representation. In the isospin formalism, the mesons ${\bar K}^0$ and $K^-$ are two isospin states of the $\bar K$ particle with the isospin of $\frac12$. Nucleon is the isospin $\frac12$ particle having two states (proton and neutron) with different projections. The total isospin is the sum of the spins of the three particles, which is independent of their other physical properties. From another side, the projections of particle isospin are determined by particle charges. It is the reason to consider "charge basis" instead of isospin "natural" basis \cite{FV20}.

According to Ref. \cite{FV20}, to separate particles sets of the $NN{\bar K}(s_{NN}=0)$ system, we take into account the difference between masses of nucleons and kaons and the charge independence breaking of nucleon-nucleon interaction. After that, the $NN{\bar K}(s_{NN}=0)$ system is represented by four configurations: $ppK^-$, $npK^-$, $np{\bar K}^0$, $nn{\bar K}^0$.
In the present work, we illustrate this separation and show the results of calculations for the $NN{\bar K}(s_{NN}=0)$ bound state.
The calculated are
mass and energy spectra of the system related to the particle configurations.

Our consideration is based on the configuration space Faddeev equations \cite{FaddeevConfigurSpace}.
For the calculations, we apply the phenomenological Akaishi-Yamazaki $N\bar K$ potential \cite{YA07} and nucleon-nucleon potential with charge symmetry breaking correction \cite{SFV19}.
The Coulomb potential is ignored in this consideration because the Coulomb correction for three-body energy may cover the effect considered. Also, we take into account that the Coulomb force violates the $AAB$ symmetry of the $np K^-$ particle configuration.


\section{Faddeev equation for  $NN\bar K$ system}
The kaonic cluster $NN\bar K$ is represented by the three-body system
with two identical particles ($AAB$ system).
The total wave function of the $AAB$ system is decomposed into the sum of the Faddeev components $U$ and $W$ corresponding to the $(AA)B$ and $A(AB)$ types of rearrangements: $\Psi =U+W-PW$, where $P$ is the permutation operator for two identical particles. In the expression for $\Psi$, the sign ''$- $'' corresponds to two identical fermions. The wave function is antisymmetrized according to two identical fermions in the system.
Each component is expressed by corresponding Jacobi coordinates.
For the system the
set of the Faddeev equations is presented by
two equations for the components $U$ and $W$ \cite{14}:
\begin{equation}
\begin{array}{l}
{(H_{0}+V_{AA}-E)U=-V_{AA}(W- PW),} \\
{(H_{0}+V_{AB}-E)W=-V_{AB}(U- PW),}
\end{array}
\label{GrindEQ__1_}
\end{equation}%
where $ H^U_{0}$ and $ H^W_{0}$
are the kinetic energy operators presented in the Jacobi coordinates for corresponding rearrangement.
In the presented work, we consider the $s$-wave approach for the $AAB$ systems. The
total angular momentum and angular momenta in the subsystem are equal to zero.
In Eq. (\ref{GrindEQ__1_}), the Faddeev component $U$ ($W$) is expressed in terms of spin and isospin spaces:
$
U={\cal U} \chi_{spin}\oplus \eta_{isospin}.
$
We consider the $NN{\bar K}$ system with the triplet isospin state of the nucleon pair $I_{NN}=1$.
The isospin basis for $NN{\bar K}$ system in the state $I=1/2$ and $I_3=1/2$ can be written using the isospin functions: $\eta_{+-+}=\eta_+(1)\eta_-(2)\eta_+(3)$, $\eta_{-++}=\eta_-(1)\eta_+(2)\eta_+(3)$, $\eta_{++-}=\eta_+(1)\eta_+(2)\eta_-(3)$.
The spin states of the $NN{\bar K}$ system is described by spin states of nucleon pair, which is the spin-singlet state.

The separation of spin/isospin variables leads to the following form of the Faddeev equations:
\begin{equation}
\label{eq:1comp}
\begin{array}{l}
(H^U_0+v^{t}_{NN}-E){\cal U}
=-v^{t}_{ NN}(-\frac12 {\cal W}^t-\frac{\sqrt{3}}2 {\cal W}^s-\frac12 p{\cal W}^t-\frac{\sqrt{3}}2 p{\cal W}^s), \\
(H^W_0+v^{s}_{\bar{K}N}-E){\cal W}^s=-v^{s}_{\bar{K}N}((-\frac{1}2 {\cal U}+ \frac{\sqrt{3}}2 p{\cal W}^t- \frac12p{\cal W}^s), \\
(H^W_0+v^{t}_{\bar{K}N}-E){\cal W}^t
=-v^{t}_{\bar{K}N}(-\frac{\sqrt{3}}2{\cal U}+\frac12 p{\cal W}^t+\frac{\sqrt{3}}2 p{\cal W}^s).
\end{array}
\end{equation}
Here, the isospin singlet and triplet ${\cal W}$ components of the wave function and corresponding isospin components of the potentials are noted by indexes $s$ and $t$. The
exchange operator $p$ acts on the particles' coordinates only. The $s$-wave spin/isospin dependent $V_{NN}$ and $V_{N{\bar K}}$ potentials are assumed for the corresponding interactions.

\section{Particle configurations}
The kaonic cluster $NN\bar K$ is represented by the three-body system
with two identical particles ($AAB$ system).
The total wave function of the $AAB$ system is decomposed into the sum of the Faddeev components $U$ and $W$ corresponding to the $(AA)B$ and $A(AB)$ types of rearrangements: $\Psi =U+W-PW$, where $P$ is the permutation operator for two identical particles. In the expression for $\Psi$, the sign ''$- $'' corresponds to two identical fermions. The wave function is antisymmetrized, according to two identical fermions in the system.
Each component is expressed by corresponding Jacobi coordinates.
For the system the
set of the Faddeev equations is presented by
two equations for the components $U$ and $W$ \cite{14}:
The $NN{\bar K}(s_{NN}=0)$ system is represented by four configurations: $ppK^-$, $npK^-$, $np{\bar K}^0$, $nn{\bar K}^0$.
The set of pair potentials and masses of particles in Eq. (\ref{eq:1comp}) definite each configuration.
For the system $ppK^-$, we have $V_{NN}=v_{pp}$ and $m_N=m_p$, $m_{\bar K}=m_{K^-}$. Where $v_{pp}$ is spin-singlet proton-proton
potential , $m_p$ is the proton mass, and $m_{K^-}$ is the mass of $K^-$. The input for the $npK^-$, $np{\bar K}^0$, $nn{\bar K}^0$ configurations is chosen by similar way using the $pp$, $pn$ and $nn$ components of spin-singlet nucleon-nucleon potential. The difference in masses of nucleons and masses of kaons are also taken into account.

To describe the spin-singlet $NN$ interaction ($I_{NN}=1$), we are applying the semi-realistic $s$-wave Malfliet-Tjon MT I-III potential with the modification from Ref. \cite{MT1}:
\begin{equation}
V_{NN}(r)=(-513.968exp(-1.55r)+1438.72exp(-3.11r))/r ,
\label{NN}
\end{equation}
where the strength (range) parameters are given in MeV (in fm$^{-1}$).
The $pp$, $np$ and $nn$ potentials are obtained by scaling the potential ({\ref{NN}}) by the factor $\gamma$ chosen in \cite{SFV19} as 0.9745,1, 0.982, respectively.

Particle configurations can be also described in the nuclear $NNN$ system as the $nnp$, $ppn$ set depending on pair potentials and mass used. The $AAB$ model with two identical particles for the $NNN$ system was proposed in Ref. \cite{SFV19}. The charge symmetry breaking for the nucleon-nucleon potential and the mass difference of proton and nucleon lead to the difference of $nnp$ and $ppn$ systems (if the Coulomb force is ignored). In particular, one can associate
the $NNN$ system with the $^3$H nucleus, when spin-singlet $nn$ potential for identical particles is used, and with the $^3$H nucleus,
when spin-singlet $pp$ potential is used. The detailed description of this model may be found in Ref. \cite{SFV19}.

We employ the $s$-wave Akaishi-Yamazaki
(AY) \cite{YA07} effective potential for $N\bar{K}$ interaction. The potential
includes the $N\bar K/\pi\Sigma$ coupled-channel dynamics into a single
channel $N \bar{K}$ interaction.
The potential is written in the form of one range Gaussian:
$
V^{s(t)}_{N{\bar K}}(r)=V_0^{s(t)}exp((-r/b)^2),
$
where the parameters are depended on pair isospin which defines the singlet $ (s)$ and triplet $(t)$ states:
$V_0^{s}=-595.0$~MeV, $V_0^{t}=-175.0$~MeV, $b=0.66$~fm.
According to Eq. (\ref{eq:1comp}), the $N \bar{K}$ potential is the same for all configurations.

We would like to note that a "particle model" which ignores isospins in the charged "$ppK^-$" kaonic system and defines pair $pK^-$, $n \bar{ K^0}$,
$p \bar{ K^0}$ potentials was considered in Ref. \cite{VFK20}. This model motivated by Ref. \cite{R16} presents the coupled $ppK^-$ and $np{\bar K^0}$ particle configurations as "$ppK^-/np{\bar K^0}$ cluster". The numerical analysis for such a two-level system had shown essential disagreement with recently reported experimental data.

The Coulomb interaction is ignored in our consideration due to its contribution, which can hide the effects related to nucleon-nucleon potentials and particle mass differences.

\section{Modeling}
\subsection{Method}
The numerical solution of the Faddeev equations  (\ref{GrindEQ__1_}) was obtained by the cluster reduction method \cite{CRM,CRM1} in  which  we used the expansion of the components $U$ and $W$ along with the basis of the eigenfunctions of  two-body  Hamiltonians of the subsystems:
\begin{equation}
U(x,y) = \sum_{i\ge 1} \phi^{U}_{i}(x)F^{U}_{i}(y), \quad
W(x,y) = \sum_{i\ge 1} \phi^{W}_{i}(x)F^{W}_{i}(y).
\label{Ex0}
\end{equation}
Here, the functions  $F^U_{i}$ and $F^W_{i}$, $i=1,2,\dots$ are describe  the relative motion of ''clusters'' in each rearrangement channel
$(NN)\bar K$ and $N(N\bar K)$, respectively. The  functions  $F^U_{i}$ ($F^W_{i}$) depend  on the 
relative  coordinate  $y$. 
The  solutions of the two-body Schr\"odinger equations form complete set of eigenfunctions in the box, 
$x \subset [0,R_x]$:
$$
(-\frac{\hbar ^{2}}{2\mu^U}\partial _{x}^{2}+V^{s_{NN}=0}_{NN}(x))\phi^U_i(x)~=~\epsilon^U_i\phi^U_i(x), \\
(-\frac{\hbar ^{2}}{2\mu^W}\partial _{x}^{2}+V_{N{\bar K}}(x))\phi^W_i(x)~=~\epsilon^W_i\phi^W_i(x),
$$
where, $\mu^U$ (and $\mu^W$) is  reduced mass of the pairs, 
$\phi^U_i(0)=\phi^U_i(R_x)=0$ ($\phi^W_i(0)=\phi^W_i(R_x)=0$),\quad $i=1,2,\dots, p$. The parameter $R_x$ is chosen to be enough large to reproduce the binding energy of $N\bar K$ subsystem in the isospin singlet channel (see also Ref. \cite{F2020}).   In the presented calculations,  $R_x$=35~fm was chosen that 
provides the convergence of the results of the three-body calculations  for $p \le 70$.

\subsection{Numerical results}

\begin{table}[h]
\caption{\label{t1} The energies and masses (in MeV) of the quasi-bound states for different particle configurations of the $NN\bar K (S_{NN}=0)$ system.}
\begin{center}
\begin{tabular}{cccccccc}
\hline
Configuration&$NN$&$m_N$& $m_{\bar K}$&$E_2$ &$E_3$ &$E_2-E_3$&$M$ \\
\hline
$ppK^-$ & $pp$ & 938.272& 493.677 &-30.259 & -45.7 &15.4& 2324.5\\
$np\bar K^0$&$np$&938.920&497.611 &-30.867&-46.7&15.8 &2328.7\\
$npK^-$ & $np$&938.920& 493.677 &-30.285&-46.1& 15.8&2325.4\\
$nn\bar K^0$&$nn$&939.565 & 497.611&-30.893&-46.5&15.6&2330.2\\ \hline
$NN\bar K$ &$np$&938.920& 495.644 &-30.576&-46.5&15.9& 2327.0\\
\hline
\end{tabular}
\end{center}
\end{table}

\begin{figure}[!th]
\centering
\resizebox{0.6\textwidth}{!}{%
\includegraphics{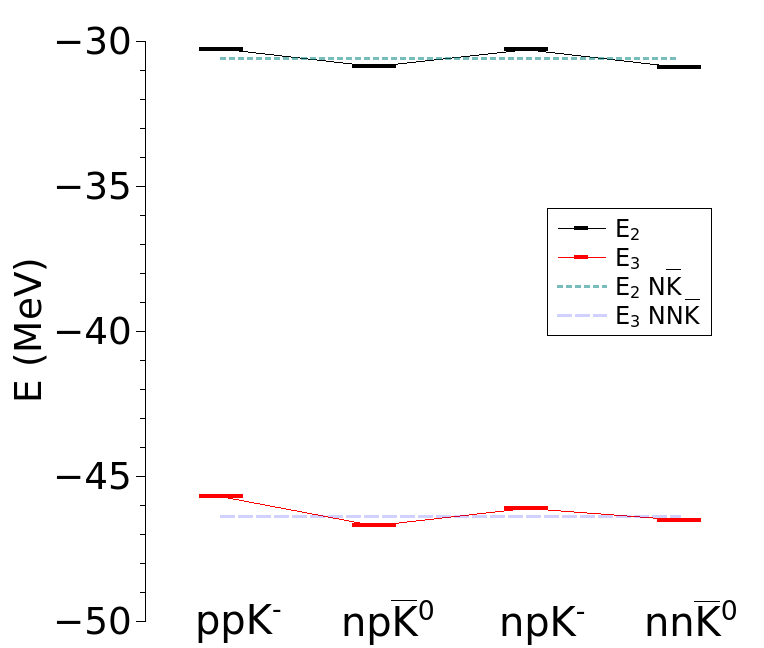}
}
\caption{ \small The two and three-body energies of the particle configurations of $NN\bar K$ system. The energy is measured from a three-body threshold. The horizontal dashed lines represent the averaged isospin model with results shown in the lower row of Tab. \ref{t1}.
}
\label{fig1}
\end{figure}
The results of the calculations for the $NN{\bar K}$ ground state energy are presented in Table \ref{t1} for the different particle sets of the $NN{\bar K}$ model. In the table, we show input parameters of each particle configuration, nucleon-nucleon potentials, and masses of the particles. For the $npK^-$ and $np{\bar K}^0$ configurations, we apply the averaged mass on nucleon to keep the $AAB$ model. For more rigorous consideration, the $ACB$ model is needed to use. Here, $E_2$ and $E_3$ are two and three-body binding energies of the $N\bar K(I=0)$ and $NN{\bar K}$ systems. The system is bound due to the attractive isospin-singlet component $(I=0)$ of the $N\bar K$ potential and weak attractive spin-singlet $NN$ potential. The last one does not form a two-body bound state. The $E_2$ is related to the experimental value for the defect mass about 27 MeV for $\Lambda$-hyperon resonance $\Lambda(1405)$ considering as the bound $N\bar K$ state \cite{170}.
The $E_2-E_3$ is the binding energy of one of the nucleons relative to a two-body threshold. The $M$ is the mass of a configuration calculated as the sum of free mass particles minus the $E_3$ (the mass defect) measuring in MeV.
Obtained results are comparable with the results of calculations performed within different approaches. For example, calculated value $|E_3|$ reported in Ref. \cite{S20} is about 46.6~MeV for phenomenological ${\bar K}N$ potential with two-pole $\Lambda(1405)$ structure (see also Ref. \cite{S2019}).

\begin{figure}[!t]
\centering
\resizebox{0.6\textwidth}{!}{%
\includegraphics{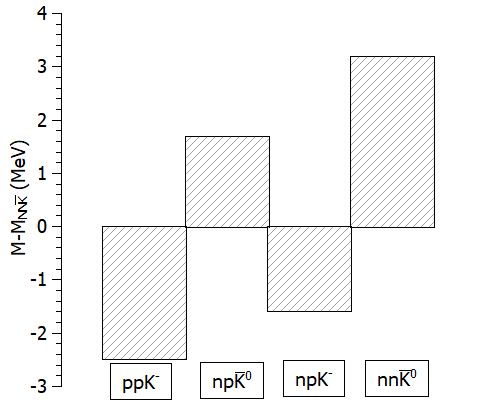}
}
\caption{\small The mass difference $M-M_{NN\bar K}$ for particle configurations in the $NN{\bar K}(s_{NN}=0)$ system.
}
\label{fig2}
\end{figure}

\begin{figure}[!ht]
\centering
\resizebox{.6\textwidth}{!}{\includegraphics{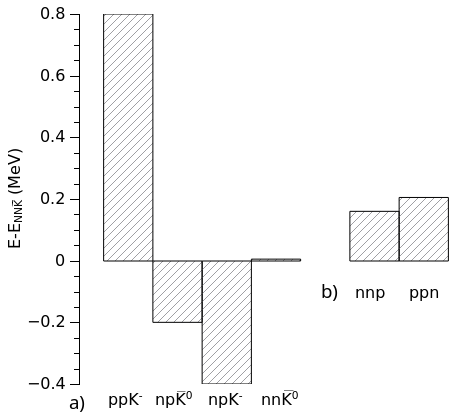}}
\caption{\small The energy difference a) $E-E_{NN\bar K}$ for the particle configurations in $NN\bar K$ system,
b) $E-E_{NNN}$ for the particle configurations in $NNN$ system.
}
\label{fig3}
\end{figure}

The graphical illustrations for the variation of the binding energy and masses for different particle configurations are presented in Figs. \ref{fig1}, \ref{fig2}, and \ref{fig3}.
One can see in Fig. \ref{fig1} that the variation in the three-body energy $E_3$ is following to the variation of the two-body energy $E_2$.
Thus, the values $E_2-E_3$ look 
similar for all particle configurations.
The $E_3$ variations are small compared to the scale defined by the $E_2-E_3$ energy.
The variations of total masses of the particle configurations are shown in Fig. \ref{fig2} relatively the total mass of the isospin model $NN{\bar K}$ with averaged nucleon and kaon masses (see Tabl. \ref{t1}). The mass differences $M-M_{NN\bar K}$ are less than 3.5 MeV depending on particle configuration. In other words, the deviation of the masses of particle configurations around the $NN{\bar K}(s_{NN}=0)$ mass is less than 0.15\%.
Similar variations for three-body energy are shown in Fig. \ref{fig3} a). The maximal energy difference $E-E_{NN\bar K}$
is related to the $ppK^-$ system and evaluated about 0.8~MeV. Thus, this derivation is the uncertainty for $NN{\bar K}$ calculations when the entry channel is not definite. The value represented by relative energy error is not so large and can be evaluated as 0.8/47, which is about 1.7\%.

In Fig. \ref{fig3} b), we illustrate our calculations using the comparison with calculated results for particle configurations in the $NNN$ system. Here, the results from Ref. \cite{SFV19} are implied, in which a phenomenological potential is used with the corrections to take into account the charge symmetry breaking in the $NN$ interaction.
The mass difference of proton and neutron has also been assumed.
In Fig. \ref{fig3}, it is seen that the effects related to the charge symmetry breaking and particle mass differences are stronger in the $NN\bar K$ system.

\section{Conclusions}
The $NN{\bar K}(s_{NN}=0)$ bound state problem was formulated by the configuration space Faddeev equation using
isospin dependent phenomenological $N\bar K$ potential. The model has taken into account nucleons/mesons mass differences and charge dependence of nucleon-nucleon potential.
The particle configurations are defined as $ ppK^-$, $np \bar {K^0}$, $nn{\bar K}^0$, $npK^-$.
The three-body calculations are performed to evaluate the mass and energy for each configuration.
In particular, the deviation of the masses around the $NN{\bar K}(s_{NN}=0)$ mass is less than 0.15\% of the $NN{\bar K}$ mass and less than 1.7\% of the $NN{\bar K}$ energy. Thus, the configuration splitting can be defined as a fine-structure effect, and the isospin $NN{\bar K}$ model is an acceptable approximation for the systems. The Coulomb interaction taken into account may correct the numerical values of the $ ppK^-$ and $npK^-$ energy splittings on 0.5~MeV\cite{O17}. The configuration splitting effect is comparable to the charge symmetry
breaking effect for $^3$H and $^3$He nuclei (about 0.6\% for energy).







\end{document}